# A look at the way we look at Complex Networks' Robustness and Resilience

*Ulisses Lacerda de Morais*

ISTA, ISCTE University Institute of Lisbon

*Luis Antunes*

GUESS, LabMAg University of Lisbon

## Abstract

This article offers a brief overview of the current research topics concerning strategies to mitigate the adverse effects of perturbations in complex networks. It addresses the issue of an unclear use of Robustness and Resilience terminology and proposes a common interdisciplinary framework for comparing strategies across different fields. It concludes with a high-level discussion of existing challenges and suggestions for future research.

# Introduction

In the Network Sciences there is substantial literature on the study of how perturbations affect the topology and functionality of complex networks. As the field itself, this is a highly interdisciplinary topic from applications varying across disciplines and scales, from genes interaction to prevention of epidemics, from major depression to financial and environmental collapse.

Despite the diversity of topics, there is a communal interest of the studies in the past few decades on strategies to ensure the networks' survivability. Moreover, the majority of the research can be further categorized under three streams of defensive strategies against perturbations: resistance, adaptability and anticipation.

Resistance strategies mainly deal with fine tuning the networks' features that minimize the propagation of the perturbations' adverse effects; be these within [5] [8] or across networks [3] [9]. Structural optimization is a recurrent theme in such studies [4, p.145] [5] [7] as it is well recognized that, depending on the type of perturbation (random or targeted), some topologies carry advantages over others [4, p.120] [7] [8] [10].

Adaptability strategies comprise fostering the characteristics that maximize recovery (or regeneration) rate. The effectiveness of these methods might be contingent on edges' rewiring costs or level of diversity among nodes as well as early-detection *ex post* [2], e.g.: through continuous monitoring [1] or inbuilt high sensitivity [12].

Anticipation strategies evidently refers to prevention via prediction or early-detection *ex ante*. The real-world applicability of this stream remains unclear as the debate over the predictability of complex networks' behavior matures [2].

Due to this subject's multidisciplinary, there is not yet a consensus about the use of terminology and typology, the literature is highly contextual and surrounded by ambiguity arising from different scientific frameworks. This article looks at this matter and proposes the adoption of a reconciliation method for the comparison of such survival strategies across disciplines.

## Robustness versus Resilience

Robustness and Resilience concepts are frequently used interchangeably or contextually within the literature, reason why these terms were intentionally avoided up to this point of the discourse. However, there is arguably much to gain by formalizing the distinction between these two terms in a generic enough fashion that could be later used to reconcile different fields' viewpoints.

In this regard, *Robustness* can be broadly defined as a network's capacity *to resist or deter the propagation of* adverse effects or loss of functionality. This capacity could be embedded in the network's topology (e.g.: scale-free networks against random failures [4, p.120] [5]), could be an innate feature of its components (e.g.: nodes homogeneity [2]), or could even be achieved through external intervention (e.g.: immunization [4, p.170]).

In contrast, *Resilience* can be generically defined as a network's capacity *to recover from or adapt to* adverse effects or loss of functionality. No distinction needed for the source of this capacity, as it could also be, for instance, due to engineered features (e.g.: autonomous monitoring layer [1] [6]) or the networks' self-organized structure (e.g.: modularity [2]).

Now it is possible to establish the link between Robustness and Resilience capacities (and measurements) to Resistance and Adaptability defensive strategies, respectively.

Although most networks will present some degree of both capacities, a clear distinction is especially relevant for the comparison of experiments and exchange of lessons learned across different disciplines.

## Suggestion of strategies comparison framework

In this regard, the several defensive strategies proposed in the complex networks body of knowledge could be defined over three axes' parameters: level of Robustness after the strategies' implementation, the analogous level of Resilience and, lastly, the elevation based on the strategies' effectiveness against perturbations (random, targeted, sustained, combined and so on).

Noticeably, for a given type of network configuration, these axes create a three dimensional space, where each strategy has a correspondent coordinate and thus, with sufficient points, a fitness landscape can be derived. It is worth adding that time is a fourth dimension that could be visualized via computational simulation to investigate the possibility of an adaptive landscape (particularly relevant for social networks). This framework excludes Anticipation strategies as their precise impact over the parameters is not currently measurable.

To illustrate this idea of interdisciplinary gains, two distinct undirected networks with 100 nodes were randomly generated[1] — a scale-free based on the Barabasi-Albert (BA) model [14] with preferential attachment power parameter was set to 2 with 4 nodes added each time step; and a random graph generated according to the Erdos-Renyi (ER) model [15] with 1/10 of edge probability.

Hypothetically, each of these networks could be the object of study by two distinct fields, say A and B respectively, each testing its own defensive strategy to increase their networks' resistance against different attack strategies aiming to weaken their networks' connectivity.

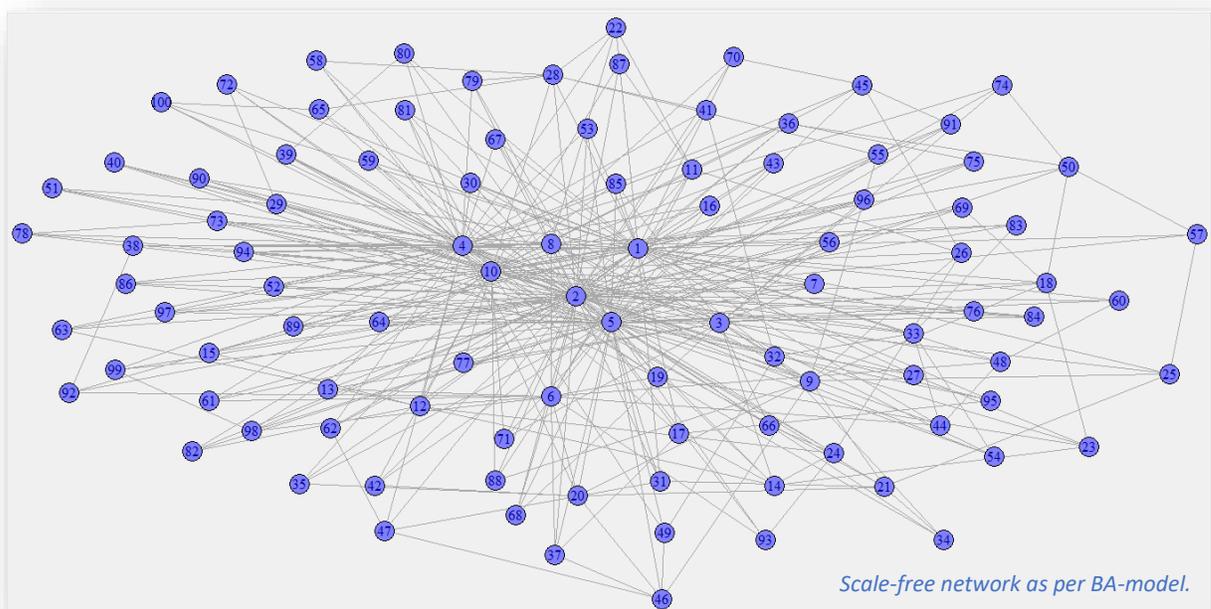

*Scale-free network as per BA-model.*

---

[1] The 'igraph R package' (http://igraph.org/r/) was used to generate the networks with RNG seed set to 101.

A look at the way we look at Complex Networks' Robustness and Resilience

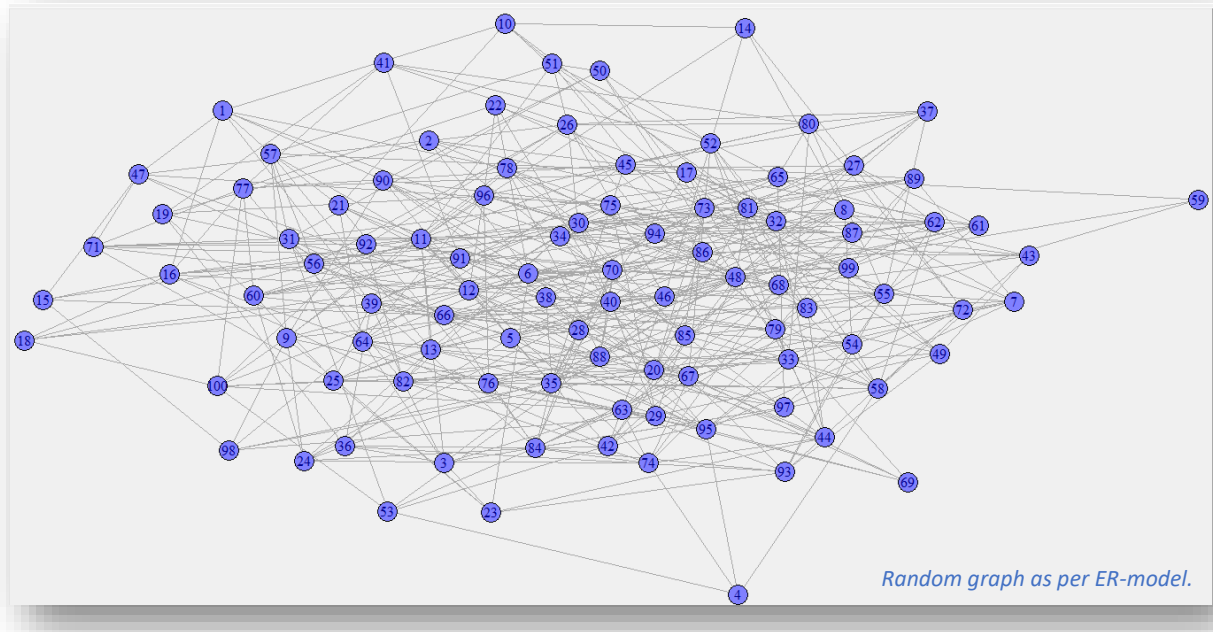
*Random graph as per ER-model.*

Field A's strategy is to randomly rewire half of its BA network's edges, while field B is creating edges between all ER network's nodes that are two hops afar from each other (neighboring).

The networks properties are shown in the table below for each of the defensive strategies:

| Original | edges | average degree | degree centrality | closeness | average path | modularity[2] (walktrap cluster) | diameter | betweenness | transitivity |
|---|---|---|---|---|---|---|---|---|---|
| BA | 390 | 7.8 | 0.6585859 | 0.5756717 | 1.98101 | 0.160572 | 3 | 0.2878281 | 0.1316674 |
| ER | 501 | 10.02 | 0.07050505 | 0.1159794 | 2.241818 | 0.2079693 | 4 | 0.02595136 | 0.0980234 |

| Rewired | edges | average degree | degree centrality | closeness | average path | Modularity[3] (walktrap cluster) | diameter | betweenness | transitivity |
|---|---|---|---|---|---|---|---|---|---|
| BA2 | 390 | 7.8 | 0.2545455 | 0.3140184 | 2.363434 | 0.2402696 | 4 | 0.128781 | 0.1271663 |
| ER2 | 501 | 10.02 | 0.09070707 | 0.1445627 | 2.233939 | 0.2419871 | 4 | 0.03502448 | 0.1104663 |

| Neighbored | edges | average degree | degree centrality | closeness | average path | modularity[3] (walktrap cluster) | diameter | betweenness | transitivity |
|---|---|---|---|---|---|---|---|---|---|
| BA3 | 4654 | 93.08 | 0.05979798 | 0.1069297 | 1.059798 | 0.01280322 | 2 | 0.0002614505 | 0.9507519 |
| ER3 | 3266 | 65.02 | 0.1987879 | 0.2529746 | 1.340202 | 0.05507451 | 2 | 0.003247331 | 0.7036844 |

---

[2] Modularity determined based on walktrap community finding algorithm which assumes that short random walks tend to stay in the same community.

Next, both networks were subjected to four node percolation sustained attack strategies[3], namely: random removal, targeted on nodes with highest betweenness centrality, targeted on nodes with the highest impact closeness (labelled "degree"), and a combination of random and targeted on betweenness (labelled "cascading").

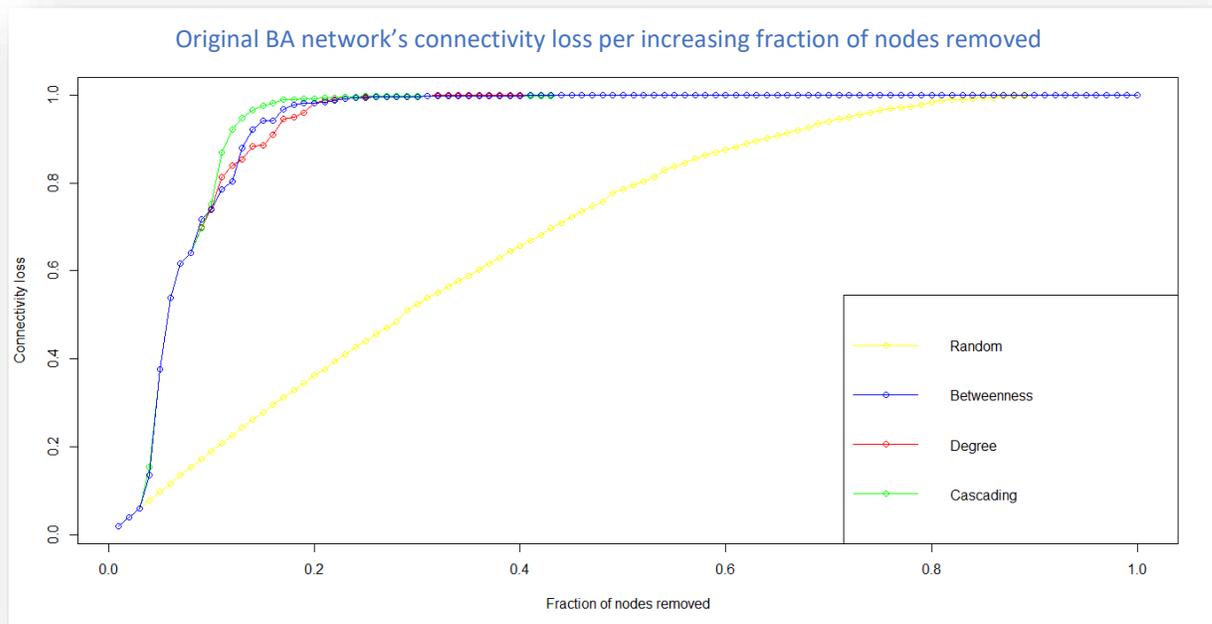

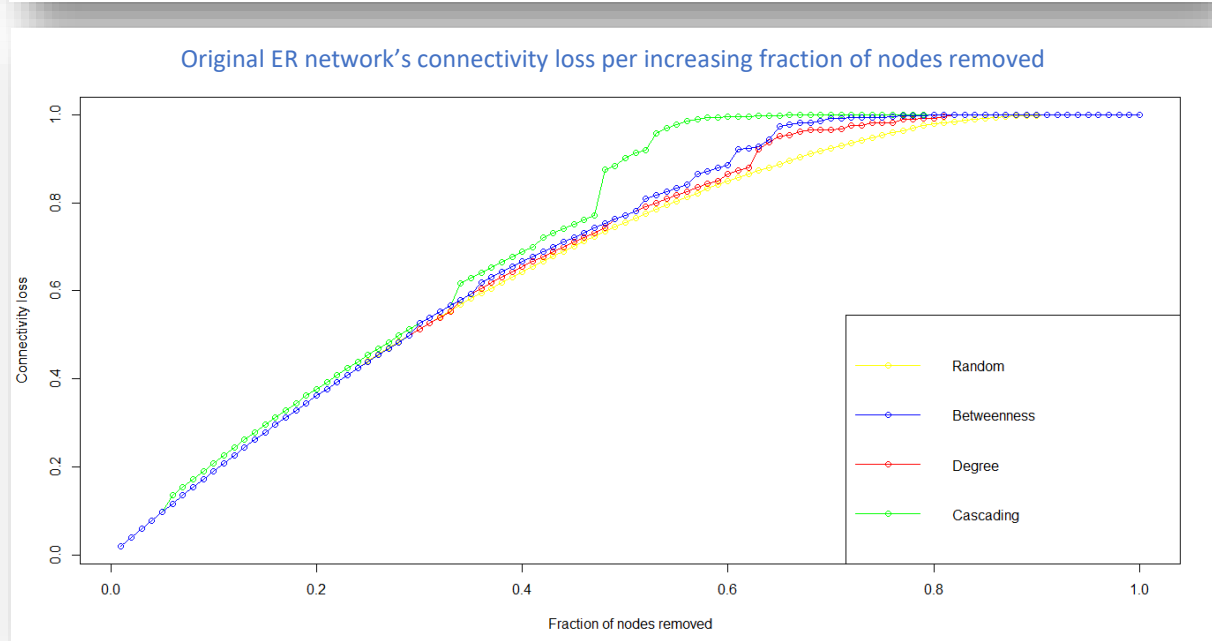

---

[3] The 'NetSwam R package' algorithm (https://CRAN.R-project.org/package=NetSwan) was used to run the robustness and resilience tests.

Field A's results after applying rewiring strategy:

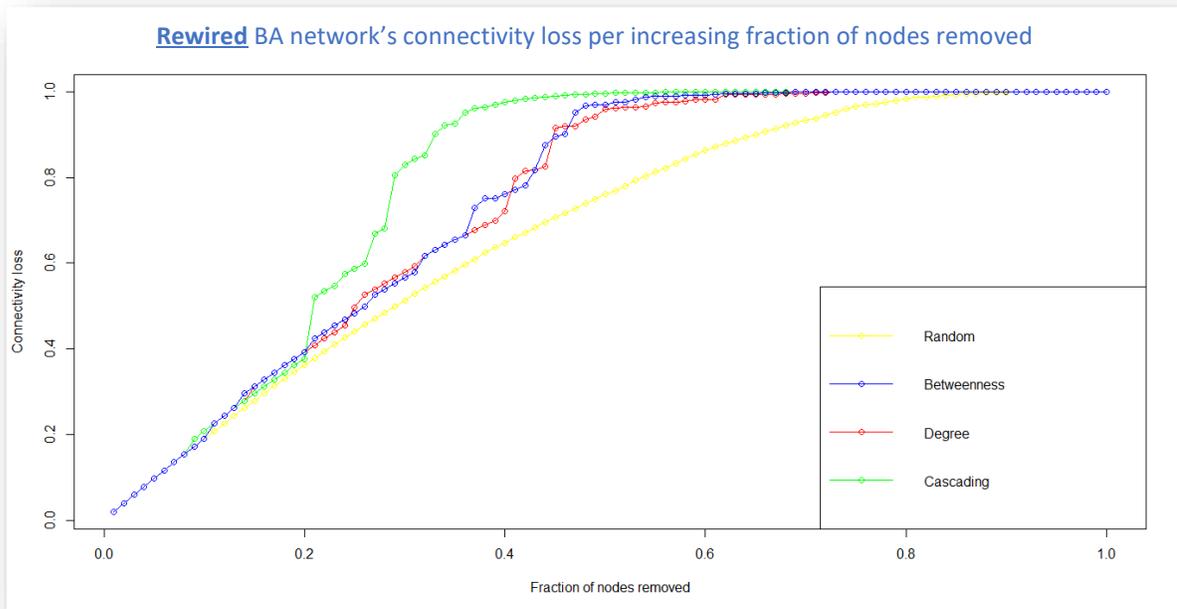

Field B's results after applying neighboring strategy:

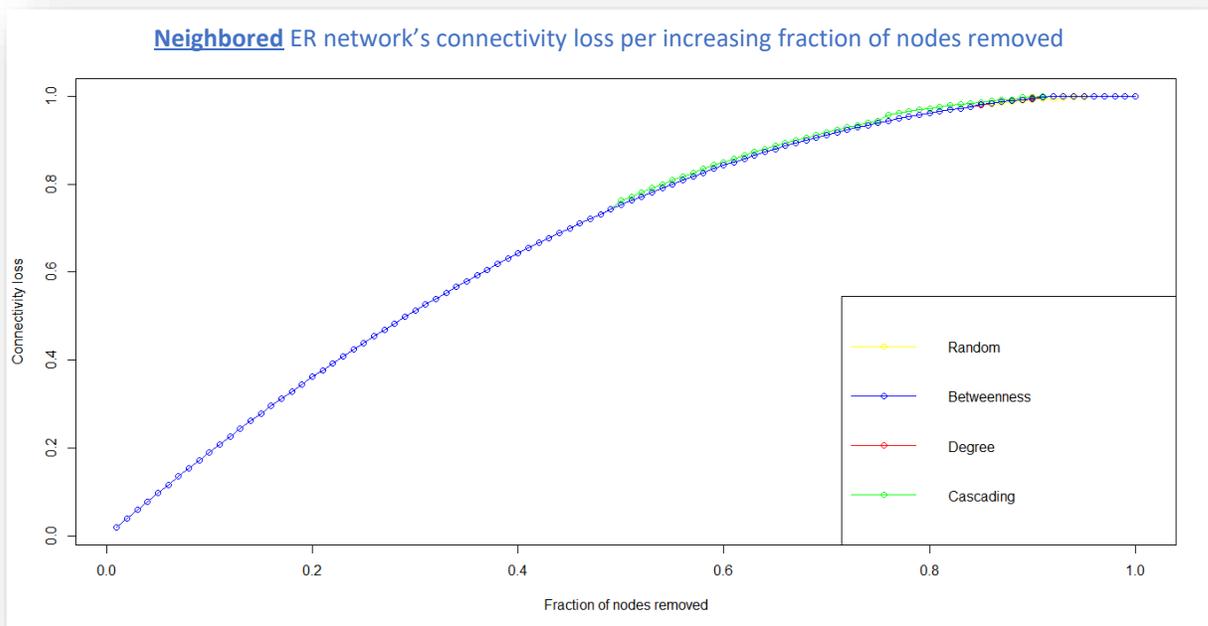

A look at the way we look at Complex Networks' Robustness and Resilience

Assuming both fields share a common method to compare their defense strategies and apply the analogous in each other's areas, and taking the decrease in the connectivity loss rate as a measure of strategies' effectiveness, then field A could benefit from applying B's strategy in its BA network. Conversely, B could learn that A's rewiring strategy is ineffective for its ER network.

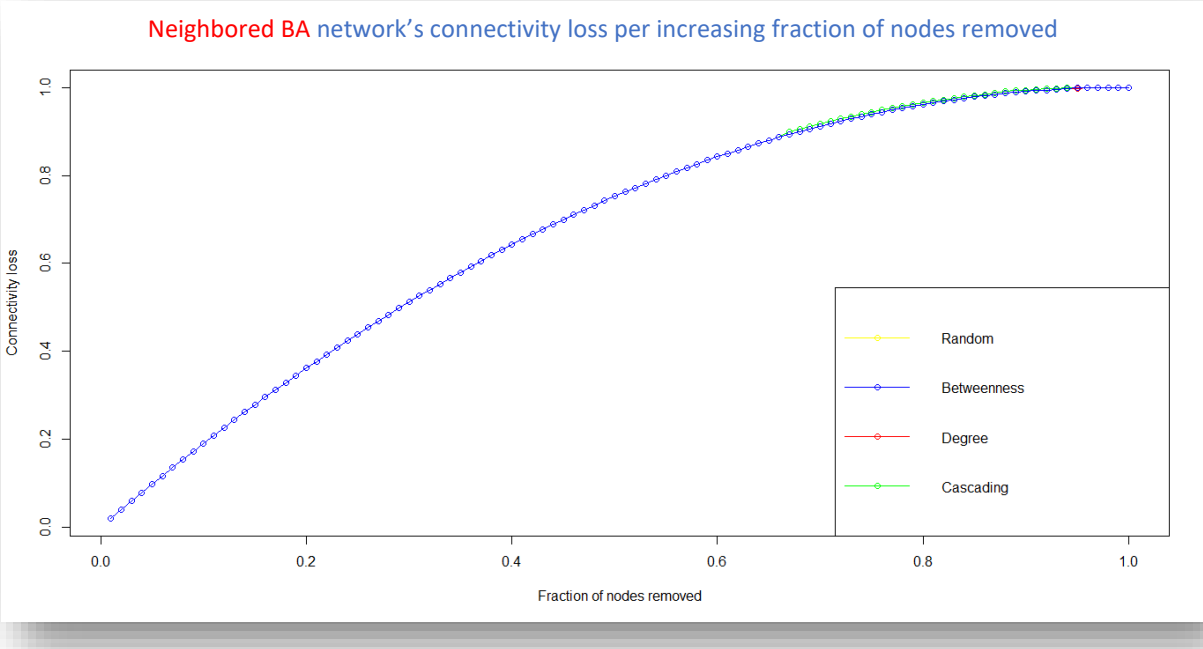

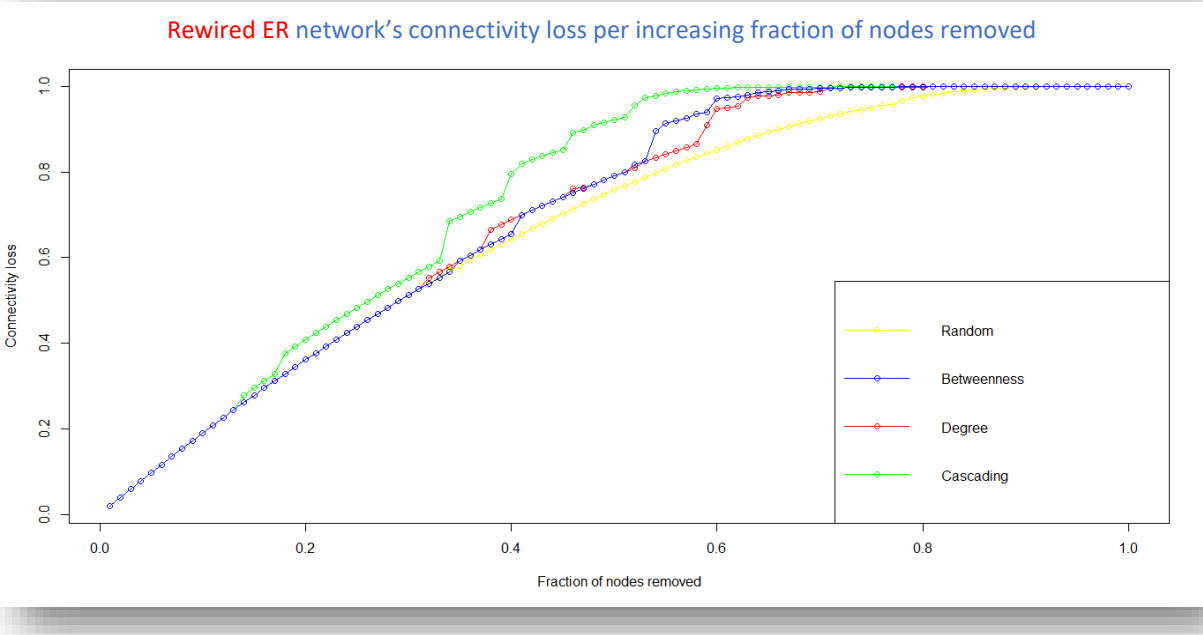

## Discussion

A bird's-eye view on the study of strategies to mitigate the adverse effects of random perturbations (failures) or targeted ones (attacks) reveals that there has been a significant advances in the body of work in the last decade, the focus is gradually shifting from simple one-layered networks with homogeneous nodes (from 60's to 80's), to multi-layered or interdependent dynamic networks with heterogeneous nodes capable to change state, learn and make local decisions on reassembling or replacement (80's onwards) [2] [3] [11].

Percolation theory and biggest connected component remain the most popular methods to test a network's resistance to disintegration [5] or cascading failures [6] [7]. However, progresses on studies of criticality [2], feedback loops [11] and adaptive capacity [12] are becoming increasingly more instrumental to improve our knowledge.

Notwithstanding, in what regards the real-world applicability and accuracy of these strategies there is much to consider. A conspicuous complication is the well observed tradeoff between Resilience and Robustness [4, p.145] [11] [12], particularly for strategies based on structure or responsiveness rate.

Structural optimization assumes that either network design or punctual intervention is possible, however reassembling costs and incomplete data create practical impediments to implementation [5] [13]. Similarly, responsiveness rate depends on the efficient flow of information across the network [12], whereas the perturbation or proximity to the critical threshold can themselves promote sluggishness on this flow [2], thus rendering the strategy ineffective.

Although a complete discussion is beyond our scope here, a fundamental inquire on drafting these strategies is: *can we prevent what we cannot predict?* This is a pertinent question for all kinds of defensive strategies. Complex networks usually contain a large degree of nonlinearity [2], also effective strategies cannot consider the network in isolation, for instance, it is demonstrated that interdependence with other networks can increase structural vulnerability [3] [9].

By definition, all research streams in this subject presume some level of prediction, from local impact or single-layered percolation (resistance strategies) to independent propagation or abundancy of resources (adaptability) to predictability itself (anticipation).

For dynamic or adaptive networks, Robustness (or Fragility) can be acquired with time, particularly in the presence of feedback loops [11]. In such cases preserving structure is suboptimal as the priority might be to maintain functionality by replacing removed nodes or edges, developing immunity after contagion, fast regeneration and etc.

Furthermore, Robustness and Resilience are not always desirable traits [12], it drastically depends on the circumstances. When the dynamics aspects are accounted for, even against random failures, robust scale-free complex networks might be susceptible to abrupt regime shifts as the entire systems could be self-organizing towards critically [2] [11].

Another factor to contemplate is that nodes may not be evenly exposed to perturbations, for example, nodes with a highest betweenness centrality could be more likely to fail due to cumulative stress [7] [9], which smudges the line between targeted and random perturbation.

In conclusion, answering the original question is not straightforward, what outlines the need for more complementarity across fields, since costly prevention cannot be incidental.

## Suggestion for future research

Building on the previous suggestion, a few interesting questions to answer in future works are, namely: what are the most effective survival strategies for complex networks? How they interplay with adversary networks? Could Resilience or Robustness capacity be built in separate layers as a control mechanism? If so, how effective is this strategy?

## References per title